\documentclass[aps,prc,twocolumn,floatfix,superscriptaddress,nofootinbib]{revtex4}

\usepackage{epsfig}
\usepackage{amsmath}
\usepackage{color}

\usepackage{graphicx}

\begin{document}
\title{Elliptic flow of thermal photons from event-by-event hydrodynamic model}
\author{Rupa Chatterjee}
\email{rupa@vecc.gov.in}
\affiliation{Department of Physics, P.O.Box 35, FI-40014 University of Jyv\"askyl\"a, Finland}
\affiliation{Variable Energy Cyclotron Centre, 1/AF, Bidhan Nagar, Kolkata-700064, India }
\author{Hannu Holopainen}
\email{holopainen@fias.uni-frankfurt.de}
\affiliation{Frankfurt Institute for Advanced Studies, Ruth-Moufang-Str. 1, D-60438 Frankfurt am Main, Germany} 
\author{Ilkka Helenius}
\email{ilkka.helenius@jyu.fi}
\affiliation{Department of Physics, P.O.Box 35, FI-40014 University of Jyv\"askyl\"a, Finland}
\affiliation{Helsinki Institute of Physics, P.O.Box 64, FI-00014 University of Helsinki, Finland}

\author{Thorsten Renk}
\email{thorsten.renk@phys.jyu.fi}
\affiliation{Department of Physics, P.O.Box 35, FI-40014 University of Jyv\"askyl\"a, Finland}
\affiliation{Helsinki Institute of Physics, P.O.Box 64, FI-00014 University of Helsinki, Finland}
\author{Kari J. Eskola}
\email{kari.eskola@phys.jyu.fi}
\affiliation{Department of Physics, P.O.Box 35, FI-40014 University of Jyv\"askyl\"a, Finland}
\affiliation{Helsinki Institute of Physics, P.O.Box 64, FI-00014 University of Helsinki, Finland}

\begin{abstract}
Elliptic flow of direct photons in relativistic heavy ion collisions is believed to be 
dominated by contribution from thermal radiation of quark gluon plasma up to $p_T$ 
$\sim 5$ GeV/$c$,  although other sources  start outshining the thermal contribution
at already smaller values of $p_T$ in the direct photon spectrum. The elliptic
flow of thermal photons from ideal hydrodynamics considering a smooth initial
density distribution under-predicts the PHENIX direct photon data from 200A GeV Au+Au 
collisions at RHIC by a large margin  in the range $1 < p_T < 5$ GeV/$c$.
However, a significant enhancement of thermal photon production 
due to fluctuations in the initial QCD matter density distributions is expected.
We show that such fluctuations result in substantially larger photon elliptic
flow for $p_T > 2.5 $ GeV/$c$ compared to a smooth initial-state-averaged density profile. 
The results from event-by-event hydrodynamics are found to be sensitive to the fluctuation 
size parameter. However, the effects of initial state fluctuations are insufficient to 
account for the discrepancy to the PHENIX data for direct photon elliptic flow. Furthermore, 
the photon $v_2$ is reduced even more when we include the NLO pQCD prompt photon component.  
We also calculate the spectra and elliptic flow of thermal photons for 2.76A TeV Pb+Pb collisions at LHC and for the 0--40\% centrality bin. Thermal photons from event-by-event hydrodynamics along with prompt photons from NLO pQCD calculations explain the ALICE preliminary direct photon data well in the region $p_T \ge 2.5$ GeV/c. Similar to RHIC, the elliptic flow results at LHC are again found to be much smaller than the ALICE preliminary $v_2$ data. 
\end{abstract}


\maketitle

\section{Introduction}
Recent fluid-dynamics simulations have shown that event-by-event (E-by-E) fluctuating
initial conditions (IC) are more realistic than smooth initial density distributions
to model the evolution of the hot and dense matter produced in relativistic heavy ion
collisions \cite{hannu,pt,scott,hannah}. Hydrodynamics with fluctuating IC
reproduces the experimental charged particle elliptic flow even for the most central
collisions at the Relativistic Heavy Ion Collider (RHIC)~\cite{hannu} which was 
underestimated by all earlier hydrodynamic calculations using smooth IC. E-by-E 
hydrodynamics also gives a better agreement of the experimental charged particle 
spectra towards higher $p_T$ by hardening the spectra~\cite{hama, hannu}, helps to 
understand the various structures observed in two-particle correlations~\cite{andrade} 
and is a necessary element in determining the shear viscosity ($\eta/s$) from simultaneous measurements of elliptic and triangular flow coefficients~\cite{eta}.

Thermal emission of photons is known to be sensitive to the initial temperature of the system where photons with large transverse momentum are emitted mostly from the hot and dense early stage of the system~\cite{phot}. Thus, they can be considered as one of the most promising probes to study fluctuations in the initial density distributions. In recent studies we have shown that E-by-E hydrodynamics with fluctuating IC enhances the production of thermal photons significantly in the region $p_T > 1$ GeV/$c$ compared to a smooth initial-state-averaged profile in an ideal hydrodynamic calculation~\cite{chre1}. This enhancement is mostly an early time effect when the radial flow is small and the 'hotspots' in the fluctuating IC produce more high $p_T$ photons than the smooth IC. 
The relative importance of IC fluctuations is found to increase for peripheral collisions and for lower beam energies~\cite{chre2}.

For a non-central collision of two spherical nuclei the overlapping zone between the
nuclei no longer remains circular but it rather takes an almond shape.
This initial spatial anisotropy of the overlapping zone is converted into
momentum space anisotropy of particle distribution via the action of azimuthally
anisotropic pressure gradients. The anisotropy is quantified by decomposing the
invariant  particle distribution in the transverse momentum plane in Fourier series as:
\begin{equation}\label{eq: v2}
 \frac{dN}{d^2p_TdY} = \frac{1}{2\pi} \frac{dN}{ p_T dp_T dY}[1+ 2\, \sum_{n=1}^{\infty} v_n (p_T) \, \rm{cos} ({\it n}\phi)] \, ,
\end{equation}
where $\phi$ is the azimuthal angle measured with respect to the reaction plane.
The most important term in the equation above is $v_2$, elliptic flow, which is related to the  almond shape mentioned above. Elliptic flow has been one of the key observables
studied at the RHIC experiments~\cite{fl2}, where
large $v_2$ values are considered as a sign of collectivity in the produced system. 

The elliptic flow of thermal photons shows interesting behavior as a function of
$p_T$ due to the interplay of the contributions from quark matter and hadronic matter
phases which dominate the flow results at different stages of the system
evolution~\cite{cfhs, sami}. The low $p_T$ part of the thermal photon elliptic flow is
dominated by the contribution from the hadronic phase whereas the high $p_T$ part
represents photons emitted from the QGP phase at the beginning of the system expansion
having small transverse and elliptic flow. As a result the thermal photon $v_2$ from
hydrodynamics is very small at large $p_T$ ($\sim 5$ GeV/$c$), where the emission
is dominated by the QGP phase. Elliptic flow rises with decreasing $p_T$ and then falls
again when $p_T$ is decreased further and the maximum is around 1.5-2.5~GeV/$c$~\cite{cfhs}.

It has been shown that the contributions from different sources of direct photons
(apart from thermal) become significant in the photon $p_T$ spectrum for
$p_T > 3$~GeV/$c$~\cite{tgfh}. However, the thermal radiation dominates the elliptic
flow of direct photons up to a much larger $p_T$ ($\sim 5$ GeV/$c$) as the $v_2$
contributions from other sources are marginal in that range~\cite{tgfh}. Prompt photons
produced in primary interactions do not exhibit any azimuthal
anisotropy and their contribution to the flow coefficient $v_2$ is zero. Photons
from fragmentation and jet conversion have a very small positive and
negative elliptic flow respectively, which tend to cancel each other~\cite{tgfh}.
Thus, the only contribution that survives in the low and intermediate $p_T$ range is
the azimuthal anisotropy of thermal photons. Thermal photon elliptic flow using
(3+1)-dimensional hydrodynamics~\cite{liu} has also been found to be quite similar
to the results obtained with a (2+1)-dimensional calculation~\cite{cfhs}.

The PHENIX Collaboration has measured a large elliptic flow of direct photons for
200A GeV Au+Au collisions at RHIC~\cite{phenix_v2}. The photon $v_2$ data shows similar
qualitative behavior as predicted by hydrodynamic calculations using smooth IC and
optical Glauber model. However quantitatively, the results from the theory 
calculations~\cite{cfhs, sami, tgfh} under-predict the data by a large margin. Similar 
results have also been observed at the LHC energy \cite {alice_v2}.

In this report we study the effect of initial state fluctuations on the elliptic
flow results of thermal photons and discuss the large difference between the  experimental 
data and results from E-by-E hydrodynamics. For recent similar investigations, discussing 
also viscous effects, see Ref.~\cite{dion}.

\section{Event-by-event hydrodynamics and direct photons}
\subsection{E-by-E hydrodynamics framework}
We use the E-by-E hydrodynamical framework  developed in ~\cite{hannu} to model
the space-time evolution of the QCD-matter. This model has been successfully used to
calculate the spectra and elliptic flow of hadrons with fluctuating IC~\cite{hannu}
as well as thermal photon spectra at RHIC and LHC energies~\cite{chre1,chre2}.
For simplicity this ideal hydrodynamical model assumes longitudinal boost invariance
and the remaining (2+1)-dimensional problem is solved numerically with the SHASTA
algorithm \cite{Boris,Zalesak}. In addition, we use the equation of state (EoS) 
from~\cite{Laine:2006cp} to close the set of equations.

To set up the initial distributions in a Monte Carlo Glauber (MCG) model the standard
two-parameter Woods-Saxon nuclear density profile is used to randomly distribute the
nucleons into the colliding nuclei. Collisions between nucleons from different nuclei
take place if the transverse distance $d$ fulfils the criterion $d^2 < \sigma_{NN}/\pi$ 
where we take the inelastic nucleon-nucleon cross section $\sigma_{NN} = 42$ and $64$~mb for RHIC and
LHC respectively.

The hydrodynamical calculation is initialized by distributing entropy density around the wounded
nucleons (sWN profile). We use a 2-dimensional Gaussian for smearing so that the initial
entropy density is 
\begin{equation}
  s(x,y) = \frac{K}{2 \pi \sigma^2} \sum_{i=1}^{\ N_{\rm WN}} \exp \Big( -\frac{(x-x_i)^2+(y-y_i)^2}{2 \sigma^2} \Big),
 \label{eq:eps}
\end{equation}
where $x_i,y_i$ are the transverse coordinates of a wounded nucleon $i$. $K$ is an
overall normalization constant used to fix the total amount of entropy and $\sigma$
is a free parameter that controls the size of the density fluctuations. We use a value
$\sigma=0.4$~fm as default \cite{hannu}, but in order to understand better the effects
from initial state fluctuations we vary the size parameter between 0.4 and 1.0~fm.
Extending studies to even smaller values of size parameter would be interesting, but
reliable calculations become numerically expensive. Since with the sWN profile the final
multiplicity (entropy) grows monotonically as a function of wounded nucleons, it is
meaningful to define centrality classes using fixed wounded nucleon ranges like was
done in~\cite{hannu,chre2}.

The initial time for the hydrodynamical calculation is taken  as in~\cite{chre2} to be $\tau_0 = 0.17 \, (0.14)$~fm/c for RHIC (LHC) motivated by EKRT minijet saturation model
\cite{hannu, Eskola:1999fc}.\footnote{For NLO pQCD systematics of $\tau_0$, see Ref.~\cite{phe}.}
The corresponding entropy normalization constants
are $K=102$~fm$^{-1}$ for RHIC and $K=250$~fm$^{-1}$ for LHC.
Freeze-out is assumed to happen on a constant temperature surface with $T_f=160$~MeV.
These choices nicely reproduce the measured $p_T$-spectra for positively charged pions
at RHIC~\cite{hannu} and LHC.

\subsection{Thermal photon emission}
The quark-gluon Compton scattering and quark-anti-quark annihilation are the leading
order processes for thermal photon production in the partonic phase. Also the
bremsstrahlung processes, which need to be taken into account in the full leading
order calculation, contribute significantly to the production~\cite{amy}. It has been shown in a very recent study that the inclusion of next to leading order (NLO) correction increases the production rate by about 20\%~\cite{nlo_jacopo} compared to the  leading order result.

In the hadronic phase $\pi$ and $\rho$ mesons contribute dominantly to the photon production
 due to the low mass of pions and the large spin iso-spin degeneracy of $\rho$ mesons~\cite{kls}.
The leading photon producing channels involving  $\pi$ and $\rho$ mesons are $\pi \pi \ \rightarrow \ \rho \gamma$, $\pi \rho \ \rightarrow \ \pi \gamma$, and $\rho \ \rightarrow \ \pi \pi \gamma$.  

As in earlier studies~\cite{chre1, chre2} we use the plasma rates $R=EdN/d^3p d^4x$ from ~\cite{amy} and hadronic rates from ~\cite{trg} (which at present can be considered as the state of the art) to calculate the spectra and elliptic flow of thermal photons from E-by-E hydrodynamics. The transition from the plasma rates to the hadronic rates is assumed to happen instantaneously at a temperature of 170 MeV.

The total thermal emission from the quark and the hadronic matter phases is obtained by integrating the rate equations over the space-time evolution of the medium,
\begin{equation}
\label{eq1}
  E\, dN/d^3p = \int d^4x\, R\Big(E^*(x),T(x)\Big),
\end{equation}
where $E^*(x)=p^{\mu}u_{\mu}(x)$. The 4-momentum of the photon is 
$p^\mu{\,=\,}(p_T \cosh Y, p_T\cos\phi,p_T\sin\phi,p_T \sinh Y)$, and the 4-velocity of the flow field is $u^\mu = \gamma_T\bigl(\cosh \eta,v_x,v_y,$ $\sinh\eta\bigr)$ with $\gamma_T = (1{-}v_T^2)^{-1/2}$, $v_T^2{\,=\,}v_x^2{+}v_y^2$. The volume element is $d^4x{\,=\,}\tau\, d\tau \, dx \, dy\, d\eta$, where $\tau{\,=\,}(t^2{-}z^2)^{1/2}$ is the longitudinal proper time and $\eta{\,=\,}\tanh^{-1}(z/t)$ is the space-time rapidity. The photon
momentum is parametrized by its rapidity $Y$, transverse momentum $p_T$, and azimuthal emission angle $\phi$. 

\subsection{Prompt photons}
We know that at sufficiently high $p_T$ the direct photon spectrum is dominated by the prompt photons originated from initial hard scatterings~\cite{tgfh,adare}. Experimentally it is not possible to separate the prompt and the thermal contributions from the direct photon spectrum. Since uncertainty arguments imply that they do not feel any medium, prompt photons are emitted isotropically and their contribution to the elliptic flow vanishes. However, their presence in the direct photon spectrum will 'wash out' the elliptic flow of thermal photons in the high $p_T$ region. In order to compare the experimental data for direct photon $v_2$ with the elliptic flow results from theory calculation, it is important to include the prompt contribution in the direct photon spectrum.

We calculate the prompt photon (direct + fragmentation) spectra in the collinear factorization framework at NLO accuracy in perturbative QCD (pQCD) using the \texttt{INCNLO}-package \cite{Aurenche:1998gv, Aversa:1988vb}. For the parton distribution functions (PDFs) we use CTEQ6.6M set \cite{Nadolsky:2008zw} with EPS09s nuclear modifications \cite{Helenius:2012wd}. The improvement in the EPS09s nuclear PDFs (nPDFs) is that due to the inclusion of impact parameter dependence, the cross sections can be calculated in different centrality classes consistently with the globally analyzed nPDFs. For the prompt photons we determine the centrality classes in terms of impact parameter intervals, which we calculate using the optical Glauber model (see~\cite{Helenius:2012wd} for detail). The parton fragmentation to photons is calculated with BFG (set II) fragmentation functions (FFs) \cite{Bourhis:1997yu} and the fragmentation process is assumed to be unmodified with respect to the vacuum fragmentation. This assumption is not expected to hold for A+A collisions in general but as the data for high $p_T$ direct photon $R_{\rm AA}$ is well reproduced with the unmodified FFs at RHIC and LHC \cite{Afanasiev:2012dg, Chatrchyan:2012vq, HeleniusEskolaPaukkunen}, the assumption is reasonable. All the relevant scales (renormalization, factorization, and fragmentation) are fixed to be equal to the photon $p_T$.

We checked using YaJEM (which is a Monte Carlo code for in-medium shower evolution)~\cite{yajem} that the medium modification enhances the fragmentation photon yield by about 25 \% compared to the result from vacuum calculation.  This in-medium modification of the fragmentation contribution does not affect the direct photon spectrum significantly as it modifies the spectrum mostly in the region $p_T < 3$ GeV/$c$ (shown later in upper panel of Figure~\ref{fig7}) where thermal radiation dominates the spectrum.

\subsection{Elliptic Flow of thermal photons}
When considering smooth initial states, the available reference plane for elliptic
flow calculation will always be the reaction plane (RP), which is defined by the
impact parameter and beam direction. However, in the experiments the impact parameter
cannot be defined. Instead in the experiments the reference plane, often called event
plane, is usually defined from the final state particles in such a way that it
maximizes the flow coefficient $v_2$.

In our case we calculate the elliptic flow with respect to the reaction plane and
in E-by-E case also with respect to participant plane (PP) (which is considered
 a good approximation for the event plane \cite{hannu}) using the relation
\begin{equation}
v_2^\gamma\{\text{PP}\}=  \langle \cos (2(\phi - \psi_{\text{PP}})) \rangle_{\text{events}} \, \,.
\end{equation}
The participant plane angle is defined as
\begin{equation}
  \psi_{\text{PP}} = \arctan \frac{ - 2\sigma_{xy} }{ \sigma_y^2 - \sigma_x^2 +
              \sqrt{ (\sigma_y^2 - \sigma_x^2)^2 + 4 \sigma_{xy}^2 }  } \, \, ,
\end{equation}
where
\begin{equation}
 \sigma_y^2  =  \langle y^2 \rangle - \langle y \rangle^2, \\ \nonumber
 \sigma_x^2  =  \langle x^2 \rangle - \langle x \rangle^2, \\ \nonumber
 \sigma_{xy}  =  \langle xy \rangle - \langle x \rangle \langle y \rangle \, . 
\end{equation}
The averaging is done over the energy density in the above equations.

\begin{figure}
\centerline{\includegraphics*[width=9.0 cm]{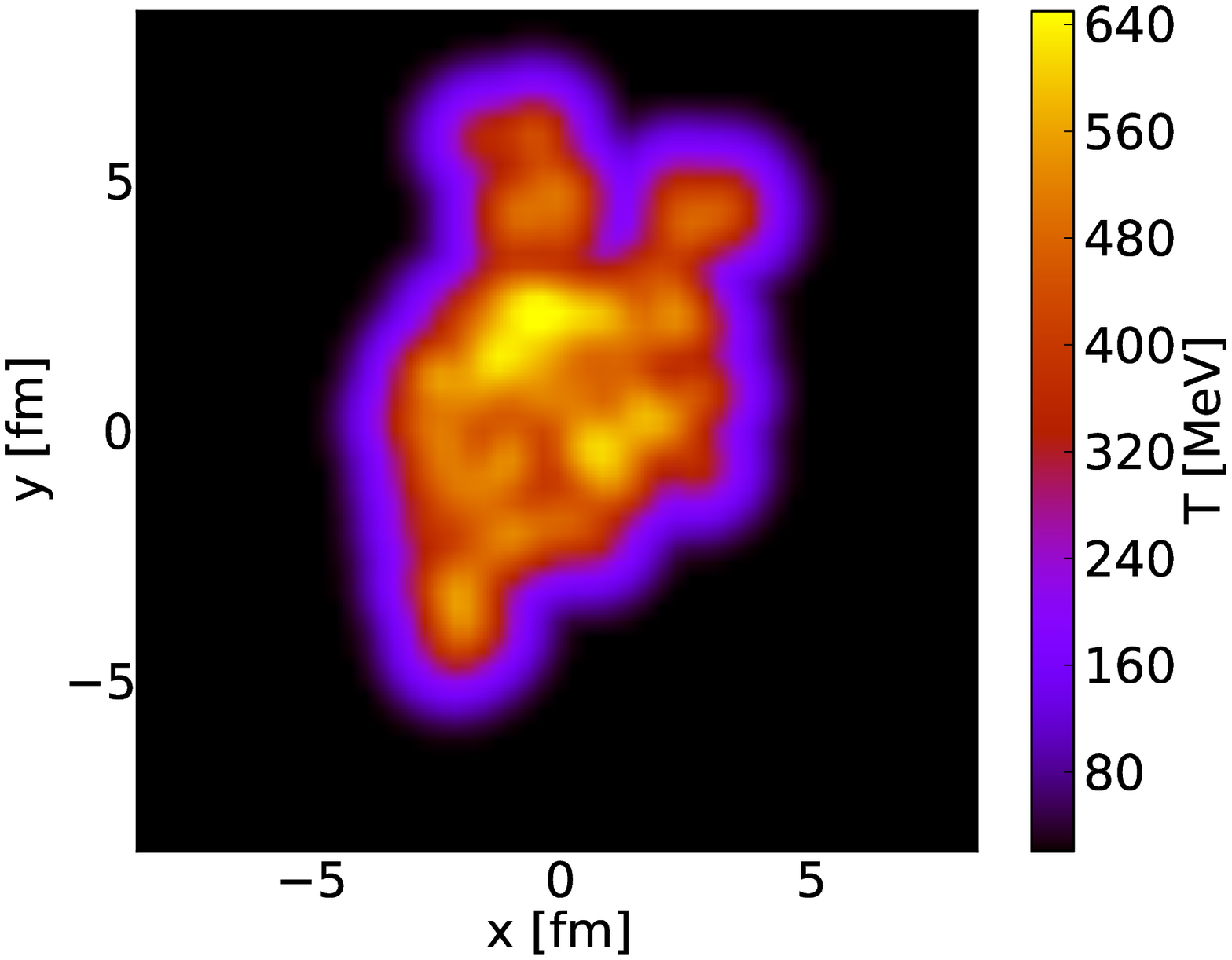}}
\centerline{\includegraphics*[width=9.0 cm]{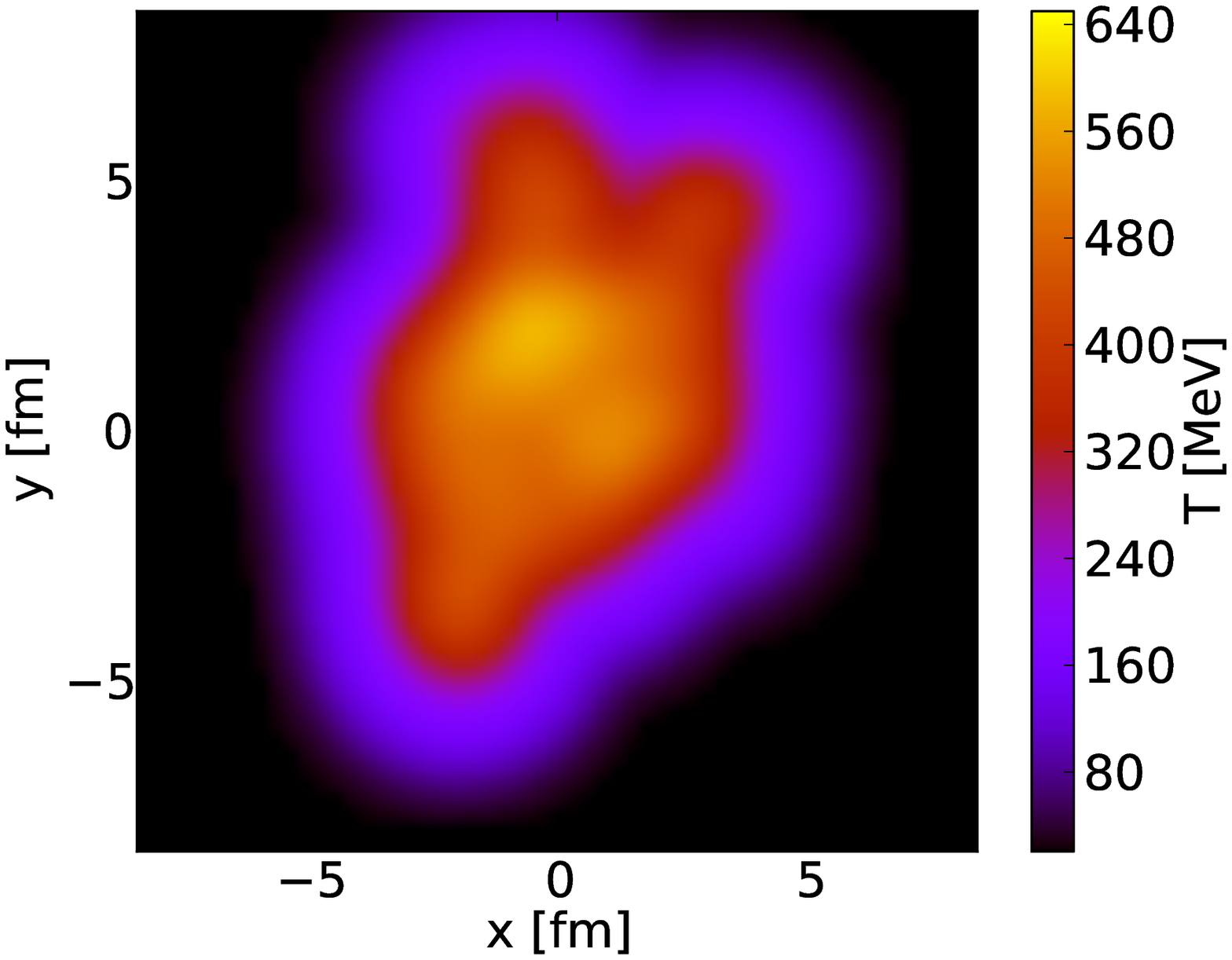}}
\caption{(Color online) Distributions of temperature in the transverse plane at time $\tau_0$ = 0.17 fm/$c$ for $\sigma$ = 0.4 (upper) and 0.8 (lower) fm and for 200A GeV Au+Au collisions at RHIC.}
\label{fig1}
\end{figure}
\begin{figure}
\centerline{\includegraphics*[width=9.0 cm]{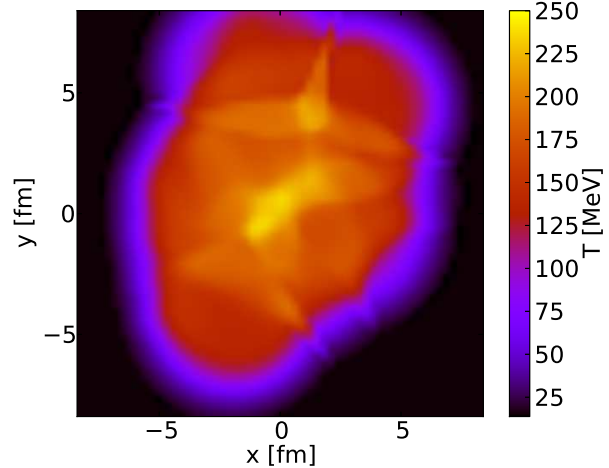}}
\centerline{\includegraphics*[width=9.0 cm]{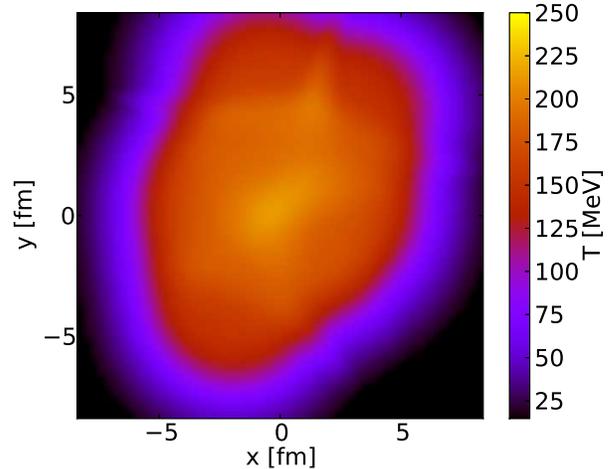}}
\caption{(Color online) Distributions of temperature in the transverse plane at time 3 fm/$c$ for $\sigma$ = 0.4 (upper) and 0.8 (lower) fm and for 200A GeV Au+Au collisions at RHIC.}
\label{fig2}
\end{figure}

\section{Results}
\subsection{Result from a single event with changing $\sigma$ values}
Thermal photons are emitted from different stages of the expanding system and thus in order to gain a better understanding  it is useful to study the time evolution of parameters like spatial anisotropy
\begin{equation}
  \epsilon_x = \frac{\int {\rm d}x {\rm d}y \, \varepsilon(x,y) (y^2 - x^2)}{\int {\rm d}x {\rm d}y \, \varepsilon(x,y) (y^2 + x^2)} \, ,
\end{equation}
momentum anisotropy
\begin{equation}
  \epsilon_p = \frac{\int {\rm d}x {\rm d}y \, (T^{xx} - T^{yy})}{\int {\rm d}x {\rm d}y \, (T^{xx} + T^{yy})} \, ,
\end{equation}
and average transverse flow velocity
\begin{equation}
  \langle v_T \rangle = \frac{\int {\rm d}x {\rm d}y \, \varepsilon(x,y) \gamma_T v_T}{\int {\rm d}x {\rm d}y \, \varepsilon(x,y) \gamma_T}
\end{equation}
from the fluctuating and smooth IC before we calculate the elliptic flow. Fluctuations
in the initial density profile can make the events in the same centrality bin 
behave differently and thus it is difficult to compare a single event with a smooth
 initial state averaged profile.
However, the initial states are smoother when we use a larger value
for the size parameter $\sigma$. Thus, we choose an event from the fluctuating IC and
change the value of $\sigma$ in that particular event from 0.4 fm to 1.0 fm
(in steps of 0.2 fm) and calculate the time evolution of transverse flow and anisotropy
parameters to see how they are affected by the smoothness of the IC.

Figures \ref{fig1} and \ref{fig2} show the temperature distributions in the
transverse plane at $\tau$ values 0.17 fm/c and 3.0 fm/c respectively for a single event from
20 -- 40\% central 200A GeV Au+Au collisions at RHIC. The corresponding values of impact parameter, $N_{\rm part}$ and $N_{\rm coll}$ for this particular event are 8.02 fm, 126 and 320 respectively and they are close to the $\langle b \rangle$, $\langle N_{\rm part}\rangle$ and $\langle N_{\rm coll}\rangle $ for 20 -- 40\% central Au+Au collisions at RHIC. The upper panels of both the figures are for $\sigma=$ 0.4 fm and the lower panels are for $\sigma=$ 0.8 fm. As expected, the hotspots are more prominent for $\sigma=$ 0.4 fm. These
hotspots in the initial profile at $\sigma=$ 0.4 fm produce more high $p_T$ photons
than the initial state averaged profile and make the spectra harder than the
smoother profile \cite{chre1}. Hydrodynamical evolution further smoothens the density
distribution, however the presence of hotspots can still be seen at $\tau = 3$~fm/$c$
for $\sigma = 0.4$~fm.

\begin{figure}
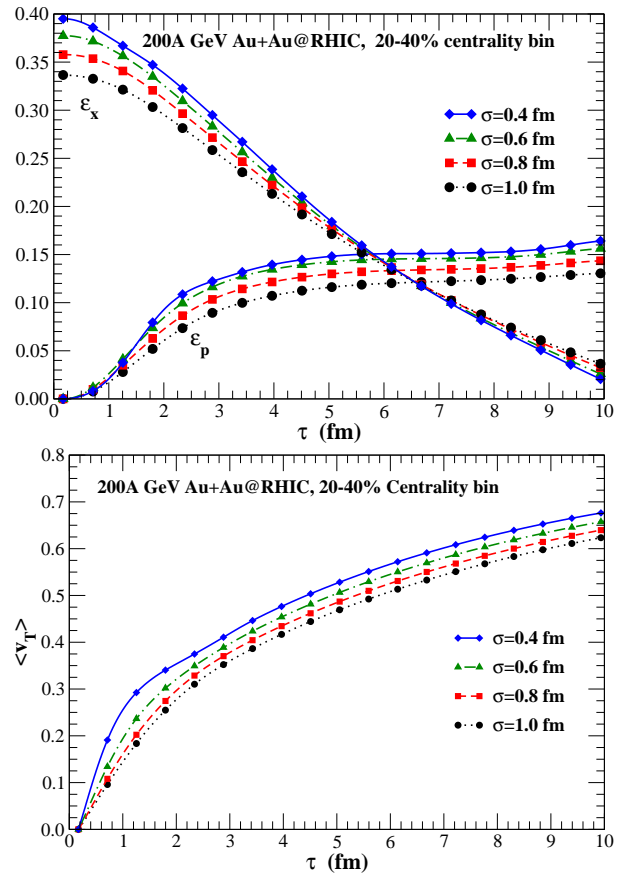
 
\centerline{\includegraphics*[width=8.0 cm]{com.eps}}
\centerline{\includegraphics*[width=8.0 cm]{vt.eps}}
\caption{(Color online) Time evolution of spatial and momentum anisotropies [upper panel] and  transverse flow velocity [lower panel] for different values of size parameter $\sigma$ for 200A GeV Au+Au collisions at RHIC and for 20--40\% centrality bin.} 
\label{fig3}
\end{figure}

The time evolution of the spatial and momentum  anisotropies
are shown in Figure~\ref{fig3} for different values of $\sigma$. The momentum anisotropy
is initially zero because there is no flow. During the evolution the pressure gradients
translate the spatial anisotropy to momentum space anisotropy. Thus $\epsilon_x$ decreases and
$\epsilon_p$ increases during the evolution. At time $\tau_0$, $\epsilon_x$ is about
15\% larger with $\sigma =$ 0.4~fm than with 1.0~fm. From the lower panel we can see that
the transverse flow develops faster with $\sigma = 0.4$~fm than 1.0~fm because 
pressure gradients in the system are larger with smaller values of the size parameter. For $\epsilon_p$
the early time behavior is not so clear, because looking carefully at Figure~\ref{fig3}
one observes that $\sigma = 0.4$~fm is not the largest scenario at early times for this particular event.
From the upper panel one sees that $\epsilon_x$ falls more rapidly with time for smaller
values of $\sigma$. Around 6.5 fm/$c$, the small and larger $\sigma$ curves intersect and
at larger times the order of the curves is opposite to the initial case. One also sees that
$\epsilon_p$ rises rapidly with time up to 2.5--3 fm/$c$ and then saturates as the spatial eccentricity becomes small with larger $\tau$.

\begin{figure}
\centerline{\includegraphics*[width=8.0 cm]{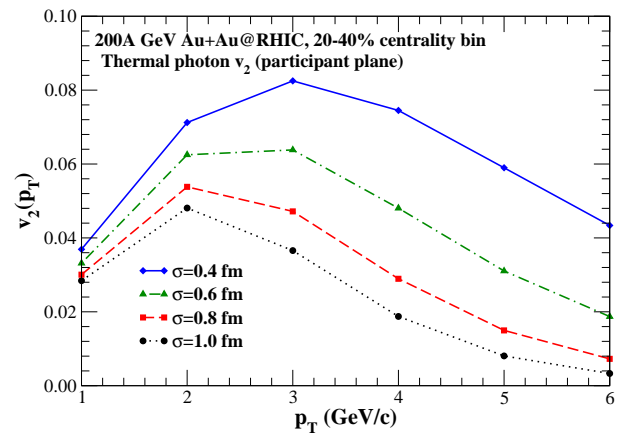}}
\caption{(Color online) Elliptic flow of thermal photons for a single event calculated with respect to the participant plane for 200A GeV Au+Au collisions at RHIC and for different values of $\sigma$.}
\label{fig4}
\end{figure}

Figure \ref{fig4} shows the elliptic flow of thermal photons calculated with
respect to the participant plane for the same event as used in Figure~\ref{fig3}
 and with different $\sigma$ values. Since the momentum anisotropy is generally
largest for the smallest size parameter, it is not surprising that the elliptic
flow is largest for the smallest $\sigma$. However, at high $p_T$ the
differences between the considered cases are larger and cannot be explained 
by looking at the momentum anisotropy alone. 

\begin{figure}
\centerline{\includegraphics*[width=8.0 cm]{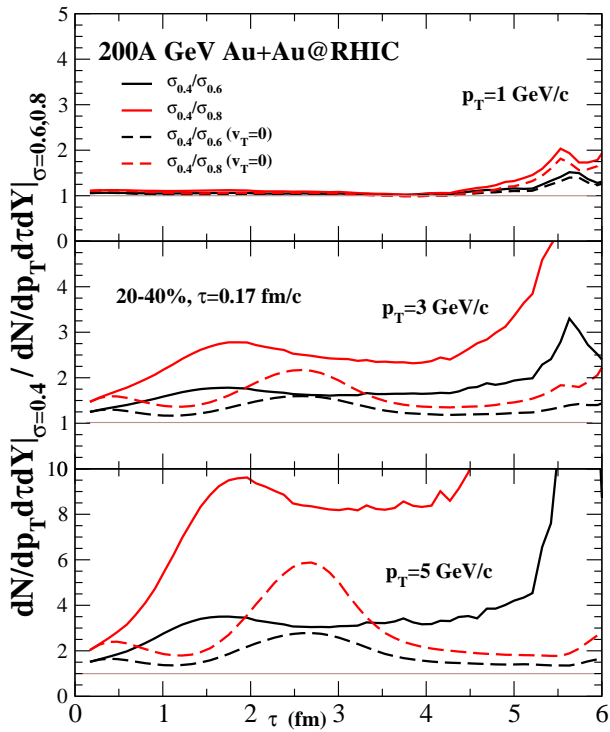}}
\caption{(Color online) The ratios of the emitted thermal photon yield as
a function of time in a single event with different size
parameters.}
\label{yield ratios}
\end{figure}

To understand better the increase in $v_2$ with smaller size parameter, 
we study the emitted photon yield as a function of time. Since we are
interested in elliptic flow, it is more meaningful to plot the ratios of
emitted yields instead of the absolute yields. These ratios are plotted in
Figure~\ref{yield ratios}. The yield ratio between $\sigma$ values 0.4 and 0.8 fm is larger than the ratio between $\sigma$ values 0.4 and 0.6 fm.  In addition we see that the  ratio is smallest when $\tau$ is very small. This means that on average the emission happens later with smallest
fluctuation size parameter.

We calculate the average emission time $\langle \tau \rangle$ at different $\sigma$ and $p_T$ values in order to understand this delay.  With $\sigma = 0.4$~fm we get $\langle \tau \rangle =
0.6$~fm/c and with $\sigma = 1.0$~fm we get $\langle \tau \rangle =
0.3$~fm/c for $p_T= 5$~GeV/c. The increase in average emission time might
seem too small to explain the huge difference in $v_2$, but for example the
momentum anisotropy is an order of magnitude larger at $\tau = 0.6$~fm/$c$ than at
$\tau = 0.3$~fm/$c$. As expected, the effect is most prominent for high $p_T$,
 and for example at $p_T=1$~GeV/$c$ the average emission time does not
change as a function of $\sigma$.

The reason for the delay in the average emission time can be due to the increased
transverse flow or due the existence of hot spots. The importance of these mechanisms
is studied by calculating the ratio of emitted yields as a function of $\tau$ keeping $v_T =$ 0 in Eq. (3) (but leaving the hydro evolution unaltered).
The results are interesting as shown by the dashed lines in Figure \ref{yield ratios}. 
At very early times ($\tau \le 0.5$ fm/$c$) the effect of hotspots is most pronounced as the 
yield ratio with  $v_T=0$ is similar to the yield ratio with non-zero $v_T$ at all $p_T$ and $\sigma$ values. For $\tau >$ 0.5 fm/$c$, transverse flow starts dominating the emission although a significant contribution from the hotspots is observed during the time period  $1.5 \le \tau \le 3.5$ fm/$c$.

\subsection{Elliptic flow from final state average at RHIC}
\begin{figure}
\centerline{\includegraphics*[width=8.0 cm]{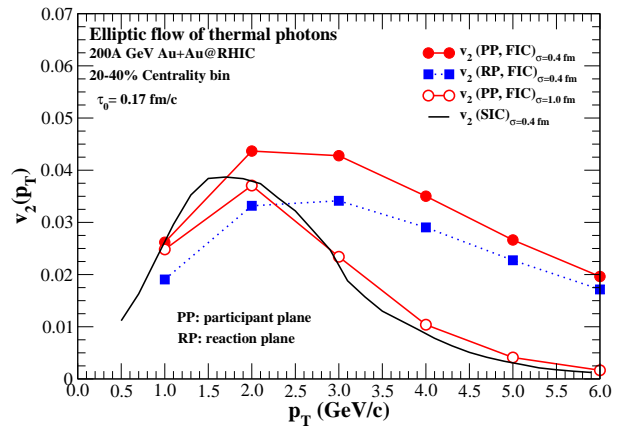}}
\caption{(Color online) Elliptic flow of thermal photons for 200A GeV Au+Au collisions at RHIC from fluctuating  and smooth IC for $\sigma=0.4$ fm. The $v_2(p_T)$ calculated with respect to the participant and reaction planes for $\sigma=0.4$ fm are shown by solid and dashed lines (closed symbols) respectively. $v_2(\rm PP)$ at $\sigma=1$ fm is shown (solid line with open symbols) for comparison. }
\label{fig5.1}
\end{figure}

Figure~\ref{fig5.1} shows the elliptic flow of thermal photons from fluctuating
 (FIC) and from smooth (SIC) initial-state-averaged IC for 200A GeV Au+Au collisions at
RHIC with $\sigma=$~0.4 and 1.0~fm. The elliptic flow of thermal photons from
the fluctuating IC is obtained by averaging over 200 random events and the
smooth initial density distribution is obtained by taking an average of 10000
fluctuating initial states~\cite{chre1}. Since in the smooth case elliptic
flow is calculated with respect to the reaction plane, in order to make a fair
comparison we compare it with reaction plane elliptic flow, $v_2(\rm{RP})$,
from fluctuating case. With $\sigma=0.4$~fm the E-by-E calculation gives
significantly larger elliptic flow for $p_T >$ 2.5 GeV/$c$ and for example at
$p_T=$ 4 GeV/$c$, the $v_2(\rm{RP})$ is about 3 times larger than the result
 from smooth IC and the difference
increases for larger values of $p_T$. However, with $\sigma=1.0$~fm the
increase in $v_2$ disappears. This behaviour was expected based on our studies
above with one single event.

Elliptic flow calculated with respect to the participant plane ($v_2(\rm{PP})$)
is even larger than the reaction plane $v_2$ in the entire $p_T$ range shown
in the figure. This behavior is similar to the hadronic case \cite{hannu} and
this happens because the initial eccentricity is larger for the
participant plane compared to the reaction plane. However, the difference between these
two reference planes seems to have some $p_T$ dependence and a detailed investigation is required to understand this 
better.

\begin{figure}
\centerline{\includegraphics*[width=8.0 cm]{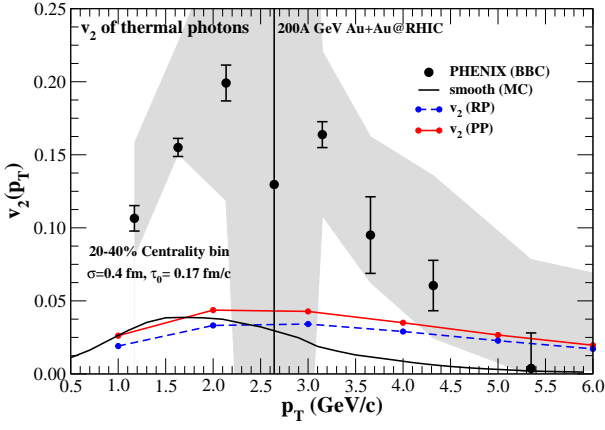}}
\caption{(Color online) Thermal photon $p_T$ spectra for 200A GeV Au+Au collisions at RHIC from fluctuating and smooth IC  and comparison with PHENIX experimental data~\cite{phenix_v2}.}
\label{fig5.3}
\end{figure}

We compare our results for thermal photon elliptic flow from the fluctuating IC
with PHENIX data~\cite{phenix_v2} in Figure~\ref{fig5.3}. We see that the
PHENIX data lie well above the results from our hydrodynamic calculations.
Fluctuations clearly bring the theory towards experiment above $p_T=2.5$~GeV/$c$,
but still below $p_T=4$~GeV/$c$ the measured values are larger than our
 calculation. Here, in discussing the thermal photons only,  we have neglected all other sources of direct photons
which will make the total photon $v_2$ from theory calculation even
smaller~\cite{sami}.

\begin{figure}
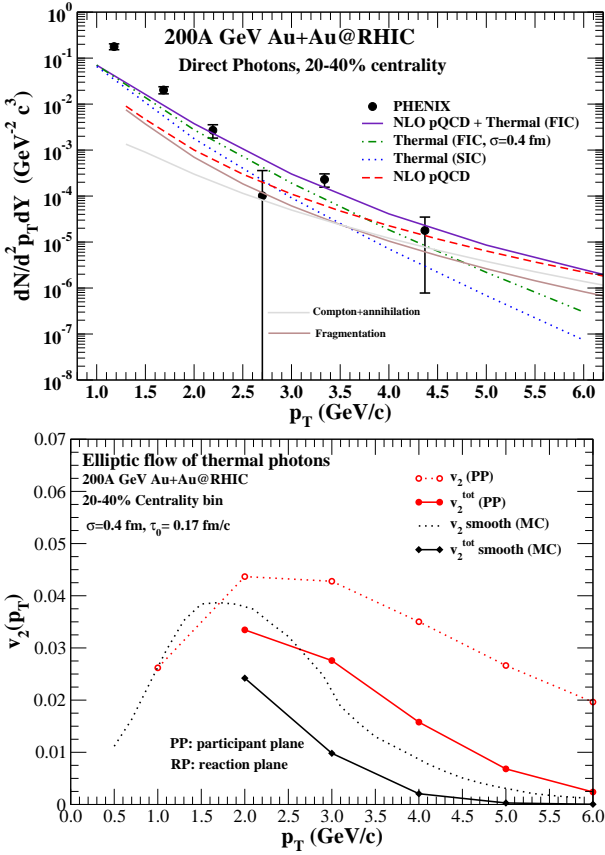

\centerline{\includegraphics*[width=8.0 cm]{prompt_rhic_spec.eps}}
\centerline{\includegraphics*[width=8.0 cm]{prompt_rhic_v2.eps}}
\caption{(Color online) [Upper panel] Direct photon spectra for 200A GeV Au+Au collisions at RHIC and for 20--40\% centrality bin~\cite{adare} along with prompt (direct+fragmentation) and thermal (fluctuating (FIC) and smooth  (SIC) initial density distributions) contributions. [Lower panel] $v_2$ with (solid) and without (dotted) the prompt photon contribution for smooth and fluctuating IC.}
\label{fig7}
\end{figure}

\subsection{Inclusion of prompt photons}
As discussed earlier, the  presence of prompt photons in the direct photon spectrum decreases the elliptic flow. The corrected spectra and elliptic flow  taking also the prompt photons into account are shown in Figure~\ref{fig7}. The PHENIX direct photon data for 200A GeV Au+Au collisions at RHIC and for 20--40\% centrality bin~\cite{adare} is compared with the prompt and thermal contributions (from smooth and fluctuating IC) in the  upper panel of Figure~\ref{fig7}. We see from the figure that the prompt photons from the NLO pQCD calculation start to dominate the direct photon spectrum for $p_T > 4$ GeV/$c$. The direct (Compton+annihilation) and the fragmentation parts of the prompt photons are shown separately.\footnote{Understanding that such a separation conceptually depends on the scale choices.} The fragmentation part dominates over the direct part for $p_T< 3.5$ GeV/$c$.  
 We see that the thermal photons from fluctuating IC ($\sigma=$ 0.4 fm) added together with the prompt photons explain the data really well in the region $p_T > 2$ GeV/$c$.  

The elliptic flow is now calculated by adding the prompt contribution using the relation
\begin{equation}
v_2 = \frac{v_2^{\rm th} . \, dN^{\rm th}\, + \, v_2^{\rm pr} . \, dN^{\rm pr}}{dN^{\rm th} \, + \, dN^{\rm pr}} 
= \frac{v_2^{\rm th} . \, dN^{\rm th}}{dN^{\rm th} \, + \, dN^{\rm pr}}  \ \ {\rm as} \ v_2^{\rm pr} \sim 0 .
\label{v2_pr}
\end{equation}
In Eq.~\ref{v2_pr} $v_2^{\rm th}$ and $v_2^{\rm pr}$ are the elliptic flow of thermal and prompt photons, respectively, and $dN^{\rm th}$ and  $dN^{\rm pr}$ are the thermal and prompt yields.
Addition of prompt contribution reduces the $v_2$ from the fluctuating IC by $\sim$25\% at $p_T=$ 2 GeV/$c$ and more than 50\% at $p_T=$ 4 GeV/$c$. The effect is larger for the $v_2$ from smooth IC than for the fluctuating IC, because fluctuations also increase the
total thermal photon yield at high-$p_T$.

\begin{figure}
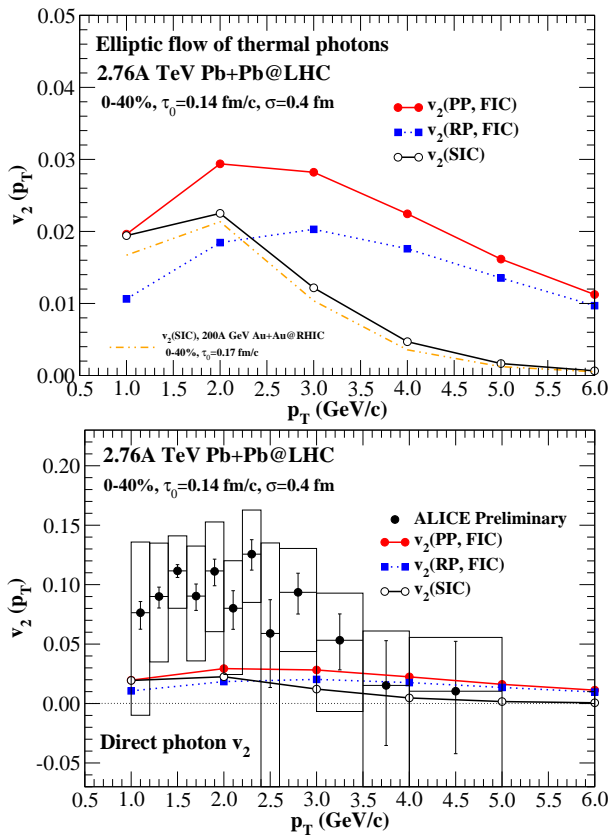

\centerline{\includegraphics*[width=8.0 cm]{v2_lhc.eps}}
\centerline{\includegraphics*[width=8.0 cm]{v2_alice.eps}}
\caption{(Color online)[Upper panel] Elliptic flow of thermal photons for 0--40\% central collisions of Pb nuclei at LHC. [Lower panel] Thermal photon elliptic flow and the ALICE preliminary direct photon $v_2$ data~\cite{alice_v2} at LHC.}
\label{fig5.5}
\end{figure}
\subsection{Elliptic flow and spectra at LHC}
The elliptic flow of thermal photons for 2.76A TeV Pb+Pb collisions at LHC and for 0--40\% centrality bin is shown in upper panel of Figure~\ref{fig5.5}. Elliptic flow results from the fluctuating IC ($v_2(\rm PP)$ and $v_2(\rm RP)$) are compared with the result obtained from a smooth initial state averaged IC. 
Similar to RHIC, fluctuations in the IC increase the elliptic flow significantly
compared to a smooth IC in the region $p_T > 2$ GeV/$c$ at LHC. 
Thermal photon $v_2$ from 200A GeV Au+Au collisions at RHIC using smooth IC is also shown for comparison. The elliptic flow at LHC is little larger than at RHIC for 0--40\% centrality bin using smooth IC. 

Our results for thermal photon elliptic flow from the fluctuating IC at LHC  are compared with the ALICE  preliminary direct photon  $v_2$ data~\cite{alice_v2} in the lower panel of Figure~\ref{fig5.5}.  As expected, the results from ideal hydrodynamic calculation are well below the experimental data for $p_T \le 3.5$   GeV/$c$.

\begin{figure}
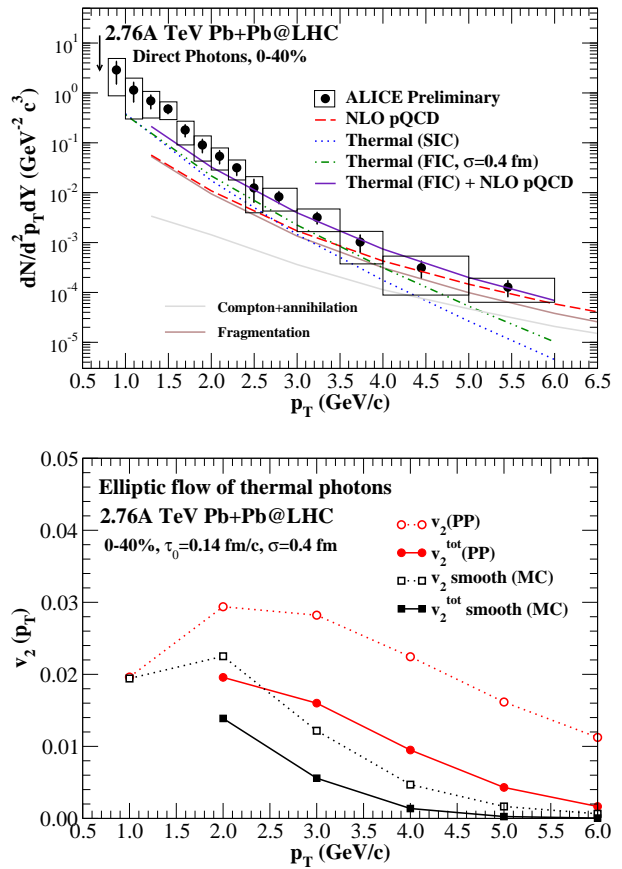

\centerline{\includegraphics*[width=8.0 cm]{spec_lhc_alice.eps}}
\centerline{\includegraphics*[width=8.0 cm]{prompt_lhc_v2.eps}}
\caption{(Color online) [Upper panel] Prompt (direct + fragmentation) and thermal (FIC and SIC) photons from 2.76A TeV Pb+Pb collisions at LHC for 0--40\% centrality bin along with ALICE preliminary direct photon data~\cite{alice_data_spec}. [Lower panel] $v_2$ with and without the prompt photon contribution for smooth and fluctuating IC.}
\label{fig10}
\end{figure}

The thermal photon $p_T$ spectra at LHC from the smooth and the fluctuating IC along with the ALICE preliminary direct photon data~\cite{alice_data_spec} are shown in the upper panel of Figure~\ref{fig10}. 
Prompt photons from NLO pQCD calculation along with the separate  direct (Compton+annihilation) and fragmentation contributions are also shown for comparison. Similar to RHIC the direct photon spectrum is dominated by the prompt photons for  $p_T > 4$ GeV/$c$ at LHC. However, unlike at RHIC  the fragmentation component at LHC is found to dominate over the direct component in the total prompt photon yield up to a very large $p_T$ ($\sim 6$ GeV/$c$). One can see that the thermal photons from the fluctuating IC added together with the prompt photons explain the direct photon spectrum well in the region $p_T > 2.5$ GeV/$c$.

Inclusion of the prompt contribution reduces the photon $v_2$ at LHC  (lower panel of Figure~\ref{fig10}) and the results are similar to the RHIC case.

\section{Summary and conclusions}
We have calculated the elliptic flow of thermal photons from an E-by-E ideal hydrodynamic
model for Au+Au collisions at RHIC and Pb+Pb collisions at the LHC. In order to
understand the physics processes underlying the photon $v_2$ better first we
 studied an individual event with different fluctuation size parameters. We saw
that a smaller $\sigma$ leads to larger momentum anisotropy and transverse flow velocity
during the hydrodynamical evolution. However, at $p_T > 2.5$~GeV/$c$ the photon
elliptic flow with a small size parameter is an order of magnitude larger than
with a large $\sigma$ and this difference cannot be understood alone from the
increase in the momentum anisotropy and the transverse flow velocity. 

To understand the increase better, we studied the photon emission as a function
of time in one single event. We see that with small size parameters the
photon emission is enhanced much more at later times compared to the early
times and thus the average emission time gets larger for smaller $\sigma$ due to 
the presence of hotspots in the IC. The elliptic flow is found to be larger for 
an E-by-E calculation at high $p_T$ region compared to the smooth IC for $\sigma = 0.4$~fm. 
In addition, $v_2(\rm PP)$ with $\sigma = 1.0$~fm is found to be very similar to the 
result from smooth IC. At small $p_T$, the E-by-E calculations produce even a bit 
smaller photon elliptic flow than the smooth IC.

As for hadrons, the calculation of elliptic flow with respect to the participant
plane gives a bit larger elliptic flow compared to the calculation with respect
to reaction plane. However, there is a $p_T$ dependence between the difference
of the PP and RP results, which should be explored more in the future. Despite
the fact that fluctuations may cause much larger $v_2$ for $p_T > 2.5$~GeV/$c$,
the enhancement is still not sufficient to explain the PHENIX measurement even
if we neglect all the other direct photon sources. We also calculated the
elliptic flow of thermal photons at LHC from smooth and fluctuating IC and compare 
our results with the ALICE preliminary data.
Similar to RHIC, fluctuations in the IC increase the elliptic flow significantly 
compared to a smooth profile for $p_T >$ 2 GeV/$c$. Also at LHC our results are 
clearly  below the measured elliptic flow by ALICE Collaboration~\cite{alice_v2}.

We also calculated prompt photons from NLO pQCD at RHIC and LHC. Thermal
photons from fluctuating IC along with prompt photons explain the PHENIX and 
the ALICE direct photon $p_T$ spectrum well in the region $p_T > 2$ GeV/$c$ and 
$p_T > 2.5$ GeV/$c$ respectively. The presence of the prompt photons in the direct photon spectrum
reduces the elliptic flow (by adding more weight in the denominator as shown in
Eq. (7)). This reduction is 20-50\% depending on the value of $p_T$ for the
fluctuating initial conditions and the reduction is even larger in the case
of smooth initial conditions, because inclusion of density fluctuations also
increases the total emitted photon yield. 

These results indicate that there is a persistent tension 
between experimental data and photons from hydrodynamics which is very large and 
not resolved by going to fluctuating hydrodynamics, but rather points 
towards an unconventional explanation.

\begin{acknowledgments} 
We gratefully acknowledge financial support by the Academy of Finland. TR and RC are supported by the Academy researcher program (project  130472) and KJE by the Academy project 133005. IH thanks the Magnus Ehrnrooth Foundation for financial support. HH is supported by the Extreme Matter Institute (EMMI). We thank ALICE Collaboration for providing us with the preliminary direct photon data. We acknowledge the computer facility of CSC -- IT Center for Science Ltd, Espoo, Finland. 

\end{acknowledgments}


\begin{thebibliography}{99}
\bibitem{hannu} H. Holopainen, H. Niemi, and K. Eskola, Phys. \ Rev. \ C {\bf 83}, 034901 (2011). 

\bibitem{pt}B. Schenke, P. Tribedy, and  R. Venugopalan,
Phys. Rev. Lett. {\bf 108}, 252301 (2012).

\bibitem{scott} J. S. Moreland, Z. Qiu, S. Bass, and U. Heinz,
Quark Matter 2012 proceedings. 

\bibitem{hannah} C. E. Coleman-Smith, H. Petersen, and R. L. Wolpert,
arXiv:1204.5774 [hep-ph].

\bibitem{hama}
Y.~Hama, T.~Kodama, and O.~Socolowski, 
Braz. J. Phys. {\bf 35}, 24 (2005).

\bibitem{andrade} R.~Andrade, F.~Grassi, Y.~Hama, T.~Kodama, and O.~Socolowski, Phys. Rev. Lett. {\bf 97}, 202302 (2006);
R.~P.~G. Andrade, F.~Grassi, Y.~Hama, T.~Kodama, and W.~L. Qian, Phys. Rev. Lett. {\bf 101}, 112301 (2008).

\bibitem{eta} B. Schenke, S. Jeon, and C. Gale,
 Phys. Rev. Lett. {\bf106}, 042301 (2011).

 


\bibitem{phot} 
  P.~V.~Ruuskanen,
  Nucl.\ Phys.\  A {\bf 544}, 169 (1992), and references therein; 
 D. K. Srivastava, \ J. \ Phys. \ G {\bf 35}, 104026 (2008).

\bibitem{chre1}
  R.~Chatterjee, H.~Holopainen, T.~Renk, and K.~J.~Eskola,
  Phys.\ Rev.\  {\bf C83}, 054908 (2011); 
  R.~Chatterjee, H.~Holopainen, T.~Renk, K.~J.~Eskola,
   J. Phys. G. Nucl. Part. Phys. {\bf 38}, 124136 (2011).

\bibitem{chre2} R. Chatterjee H. Holopainen, T. Renk, and K. J. Eskola,
Phys. \ Rev. C {\bf 85},  064910 (2012); R. Chatterjee H. Holopainen, T. Renk, and K. J. Eskola, arXiv:1207.6917.


\bibitem{fl2} C.~Adler {\it et al.} [STAR Collaboration], Phys.\ Rev.
\ Lett. {\bf 87}, 182301 (2001); ibid {\bf 89}, 132301 (2002); ibid
{\bf 90}, 032301 (2003); S.~S.~Adler {\it et al.} [PHENIX Collaboration],
Phys.\ Rev. \ Lett. {\bf 91}, 182301 (2003).



\bibitem{cfhs} R. Chatterjee, E. Frodermann, U. W. Heinz, and D. K. Srivastava, Phys. \ Rev. \ Lett. {\bf 96}, 202302  (2006);  R. Chatterjee and D. K. Srivastava, Phys. \ Rev. \ C {\bf 79}, 021901(R) (2009); U.~W.~Heinz, R.~Chatterjee, E.~S.~Frodermann, C.~Gale, and D.~K.~Srivastava,
  Nucl.\ Phys.\ A {\bf 783}, 379 (2007).


\bibitem{sami} H. Holopainen, S. Rasanen, K. J. Eskola,	
Phys. \ Rev. \ C {\bf84}, 064903 (2011).

\bibitem{tgfh} S. Turbide, C. Gale, E. Frodermann, and U. W. Heinz,
 Phys. Rev. C {\bf 77}, 024909 (2008).

 
\bibitem{liu} F. M. Liu, T. Hirano, K. Werner, and Y. Zhu, Phys. \ Rev. \ C {\bf 80}, 034905 (2009).

\bibitem{phenix_v2} A.~Adare {\it et al.}  [PHENIX Collaboration], Phys. Rev. Lett. {\bf 109}, 122302 (2012).


\bibitem{alice_v2} D. Lohner for the ALICE Collaboration, arXiv:1212.3995.


\bibitem{dion} 
  M.~Dion, J.~-F.~Paquet, B.~Schenke, C.~Young, S.~Jeon and C.~Gale,
  Phys.\ Rev.\ C {\bf 84}, 064901 (2011);  C.~Gale,
  arXiv:1208.2289 [hep-ph].


\bibitem{Zalesak}
S.~T. Zalesak,
J. Comput. Phys. {\bf A31}, 335 (1979).


\bibitem{Boris}
J.~P. Boris and D.~L. Book,
J. Comput. Phys. {\bf A11}, 38 (1973).

\bibitem{Laine:2006cp}
M.~Laine and Y.~Schroder,
Phys. Rev. {\bf D73}, 085009 (2006).




\bibitem{Eskola:1999fc}
K.~J. Eskola, K.~Kajantie, P.~V. Ruuskanen, and K.~Tuominen,
Nucl. Phys. {\bf B570}, 379 (2000).


\bibitem{phe}
R. Paatelainen, H. Holopainen, and K.~J. Eskola, Phys. Rev. C {\bf 87}, 044904 (2013).



\bibitem{amy} P.~Arnold, G.~D.~Moore, and L.~G.~Yaffe, JHEP {\bf 0112}, 009
(2001).

\bibitem{nlo_jacopo} J. Ghiglieri, J. Hong, A. Kurkela, E. Lu, G. D. Moore, and
D. Teaney, arXiv:1302:5970.


\bibitem{kls} J.~I.~Kapusta, P.~Lichard and D.~Seibert,
  Phys.\ Rev.\  D {\bf 44}, 2774 (1991)
  [Erratum-ibid.\  D {\bf 47}, 4171 (1993)].

\bibitem{trg} S.~Turbide, R.~Rapp, and C.~Gale, \ Phys. \ Rev. \ C {\bf 69},
 014903 (2004).

\bibitem{adare}
  A.~Adare {\it et al.}  [PHENIX Collaboration],
  Phys.\ Rev.\ Lett.\  {\bf 104}  132301 (2010).

\bibitem{Aurenche:1998gv}
  P.~Aurenche, M.~Fontannaz, J.~P.~.Guillet, B.~A.~Kniehl, E.~Pilon and M.~Werlen,
  Eur.\ Phys.\ J.\ C {\bf 9}, 107 (1999).

\bibitem{Aversa:1988vb}
F.~Aversa, P.~Chiappetta, M.~Greco and J.~P.~Guillet,
  Nucl.\ Phys.\ B {\bf 327} (1989) 105.


\bibitem{Nadolsky:2008zw}
  P.~M.~Nadolsky, H.~-L.~Lai, Q.~-H.~Cao, J.~Huston, J.~Pumplin, D.~Stump, W.~-K.~Tung and C.~-P.~Yuan,
  Phys.\ Rev.\ D {\bf 78}, 013004 (2008).

\bibitem{Helenius:2012wd}
  I.~Helenius, K.~J.~Eskola, H.~Honkanen and C.~A.~Salgado,
  JHEP {\bf 1207}, 073 (2012).

\bibitem{Bourhis:1997yu}
  L.~Bourhis, M.~Fontannaz and J.~P.~Guillet,
  Eur.\ Phys.\ J.\ C {\bf 2}, 529 (1998).

\bibitem{Afanasiev:2012dg}
  S.~Afanasiev {\it et al.}  [PHENIX Collaboration],
  Phys.\ Rev.\ Lett.\  {\bf 109}, 152302 (2012).

\bibitem{Chatrchyan:2012vq}
  S.~Chatrchyan {\it et al.}  [CMS Collaboration],
  Phys.\ Lett.\ B {\bf 710}, 256 (2012).

\bibitem{HeleniusEskolaPaukkunen}
  I.~Helenius, K.~J.~Eskola and H.~Paukkunen,
  JHEP {\bf 1305}, 030 (2013).


\bibitem{yajem} T. Renk, Int. \ J. \ Mod. \ Phys. {\bf E20} 1594 (2011).

\bibitem{alice_data_spec} M. Wilde for the ALICE Collaboration, arXiv:1210.5958.









\end{thebibliography}
\end{document}